\documentclass[11pt,twoside,onecolumn]{article}
\usepackage[]{latexsym,amsmath,amssymb}
\usepackage[]{epsfig}
\newcommand{\be}{\begin{eqnarray}}
\newcommand{\ee}{\end{eqnarray}}

\title{\bf Hawking temperature for various kinds of black holes
from Heisenberg uncertainty principle}
\author{Fabio Scardigli\thanks{Address for the correspondence:
via Europa 20 - 20097 S.Donato - Milano, Italy. E-mail: scardigli@fisica.ist.utl.pt}
\\
\\
{\em CENTRA, Departamento de Fisica, Instituto Superior Tecnico}\\
{\em Av. Rovisco Pais 1, 1049-001 Lisboa, Portugal}}
\date{}
\begin{document}
\maketitle
\begin{abstract}
Hawking temperature is computed for a large class of black holes (with spherical,
toroidal and hyperboloidal topologies) using only laws of classical physics plus
the "classical" Heisenberg Uncertainty Principle. This principle is shown to be
fully sufficient to get the result, and there is no need to this scope of a
Generalized Uncertainty Principle.

\par
\null
\end{abstract}
\raggedbottom
\setcounter{page}{1}
\section{Introduction}
\setcounter{equation}{0}
The Hawking temperature has been heuristically derived for various kinds of black holes
by using several different versions of the uncertainty principle. For the Schwarzschild
black hole, the "classical" Heisenberg principle alone seems to be sufficient to
obtain the Hawking formula (see, for example, Ref. \cite{Scardigli}), even if some
small adjustments of the numerical constants are sometime used \cite{Santiago}.
Recently, in Ref. \cite{Cavaglia}, the Hawking temperature for the (Anti) de Sitter
black hole has been computed with a generalized version of the uncertainty principle.
In this derivation some numerical constants are chosen in a somehow arbitrary way,
and a particular version of the generalized uncertainty principle (GUP), namely
\be
\Delta x \Delta p \geq \hbar \left(1 + \beta^2 \frac{\Delta x^2}{l_p^2}\right),
\ee
(where $\beta$ is an arbitrary constant and $l_p$ is the Planck length)
is claimed to be necessary to get the exact form of the
Hawking temperature in the (Anti) de Sitter case. From this
claim, the authors speculate on some connection between the fact that
the Anti de Sitter black hole thermodynamics admits a holographic
description in terms of a dual conformal quantum field theory, and the
fact that the GUP finds its natural collocation in string theory and in
non commutative geometry.\\
The aim of the present paper is, on the contrary, to show that the local Hawking
temperature of a quite large class of black holes (Schwarzschild,
Reissner-Nordstr\"{o}m, (Anti) de Sitter, with the various topologies, spherical,
toroidal and hyperboloidal) can be obtained directly from the "classical" Heisenberg
principle casted in the form $\Delta E \Delta x \simeq \hbar c /2$ and from the
effective (newtonian) potential approximating the metric, without need of any form
of Generalized Uncertainty Principle. The present derivation does not contain any
adjustable parameter and, although heuristic, gives us a final result in very good
agreement, even numerical, with the exact formulae worked out from the quantum field
theory on curved space-times. This result in turn can be seen as a further evidence
of the purely kinematic nature of the Hawking effect \cite{Visser} and of its
independence from the dynamical Einstein equations.

\section{Effective potential from the metric}
\setcounter{equation}{0}
We consider here space-times with a metric that locally has the form
\be
ds^2 \,=\, -F(r)c^2dt^2 \,+\, F(r)^{-1}dr^2 \,+\, r^2d\Omega_k^2 \,=
\, g_{\mu\nu}dx^\mu dx^\nu
\label{mo}
\ee
where
\be
F(r) = k -\frac{2GM}{c^2 \,r} + \frac{GQ^2}{c^4 \,r^2} + \lambda \,r^2
\label{m}
\ee
and the time-like coordinate is chosen as $x^0=ct$. The parameters $M$ (mass),
$Q$ (electric charge), $\lambda$ (cosmological constant, up to a factor) are
real and continuous.
$\lambda<0$ corresponds to a de Sitter space-time, $\lambda>0$ to an Anti de
Sitter space-time. The discrete parameter $k$ takes the values $1,\,0,\,-1$ and
$d\Omega_k^2$ is the metric on a two dimensional surface
$\Sigma_k$ of constant Gaussian curvature $k$. In local coordinates
$(\theta, \phi)$ on $\Sigma_k$ we have
\be
d\Omega_k^2 =
\left\{ \begin{array}{lll}
d\theta^2+\sin^2\theta \,d\phi^2, & k=1, & {\rm spherical},
\\
\\
d\theta^2+\theta^2 \,d\phi^2, & k=0, & {\rm toroidal},
\\
\\
d\theta^2+\sinh^2\theta \,d\phi^2, & k=-1, & {\rm hyperboloidal}.
\end{array}
\right.
\ee
The vector $\partial/\partial t$ is a Killing vector, time-like for $F>0$
and space-like for $F<0$. Incidentally, the metric (\ref{m}) solves the
Einstein-Maxwell equations with a cosmological constant $\Lambda=-3\lambda$.
However, we shall see that this fact does not enter in the derivation of the
Hawking temperature.
An expression of the effective potential may be obtained considering the equation of
motion for a neutral particle of negligible mass in a gravitational field
$g_{\mu\nu}$, given by the geodesic
\be
\frac{d^2x^\lambda}{ds^2} + \Gamma_{\mu\nu}^{\lambda}\frac{dx^\mu}{ds}\frac{dx^\nu}{ds}=0
\label{geo}
\ee
where, as usual, $\Gamma_{\mu\nu}^{\lambda}=\frac{1}{2}g^{\lambda\sigma}
(g_{\mu\sigma,\nu}+g_{\nu\sigma,\mu}-g_{\mu\nu,\sigma})$.\\
If we suppose that the particle is moving $slowly$, in a $stationary$ and $weak$
gravitational field, then we can follow well known steps, described for example
in Ref.~\cite{weinberg}, and we easily arrive to define the effective (newtonian)
potential
\be
V = -\frac{c^2}{2}\, h_{00} + C,
\label{V}
\ee
where $C$ is a constant, and where
(since the field is $weak$) the metric is supposed to be close to the
flat metric
\be
g_{\mu\nu}=\eta_{\mu\nu} + h_{\mu\nu}
\label{wf}
\ee
with $|h_{\mu\nu}| \ll 1$ and $\eta_{\mu\nu}\equiv (-k,1,1,1)$,
with $k=1,\,0,\,-1$.\\
In case $V \rightarrow 0$ and
$h_{00} \rightarrow 0$ for $r \rightarrow \infty$, then $C \,= \,0$.\\
We can now express $V$ in terms of the metric (\ref{m}). In fact
from (\ref{mo}) we have
\be
g_{00}=-F(r)
\ee
and from (\ref{wf}), when $|h_{00}| \ll 1$, we can write
\be
g_{00}=\eta_{00}+h_{00}=-k+h_{00}.
\ee
Therefore using (\ref{V}) we have
\be
V = \frac{c^2}{2}\,\,(F-k) + C.
\label{vf}
\ee
which is the effective potential generated by the metric (\ref{mo}).
\section{Horizons}
\setcounter{equation}{0}
In this section we consider some general properties of the (Killing)
horizons of the metric (\ref{mo})(\ref{m}).
As usual \cite{43louko} the horizons are located at the positive
zeros of the function $F(r)$. They are coordinate singularities on
null hyper-surfaces. The vector $\partial / \partial t$ is a globally
defined Killing vector, time-like in the regions $F>0$, space-like in
the regions $F<0$ and null on the hyper-surfaces with $F=0$. The regions
with $F>0$ are therefore static, and the hyper-surfaces $F=0$ are Killing
horizons \footnote{
For a complete analysis of the causal structure and the Penrose diagrams
of these zeros/horizons we refer the reader to Ref.~\cite{71030louko}, for
the case $k=1$, and to Ref.~\cite{louko}, for the cases $k=0$, $k=-1$}.\\
We are interested here in describing some behaviours of the zeros
of $F(r)$ for various values of the parameters $M$, $Q$, $\lambda$, and
especially what are the regions where the term
$h_{00}$ becomes small, as required by the approximation used to obtain
the effective potential $V$.\\
In units where $G=c=1$, the function $F(r)$ can be rewritten as
\be
F(r) = k -\frac{2M}{r} + \frac{Q^2}{r^2} + \lambda \,r^2.
\ee
We restrict our analysis to the case $M>0$, on the grounds of physical plausibility. \\
\\
For example, for $k=1$ and $\lambda >0$ (Anti de Sitter space-time) a straightforward
analysis reveals that for
$\Delta:=M^2-Q^2\leq 0$ there are no zeros of $F(r)$, therefore we do not
have horizons. For
$\Delta:=M^2-Q^2 > 0$ we can still have no horizons, or two coinciding
horizons, or even two different positive horizons. It is easy to show that
it is impossible to have one single horizon (because $Q^2>0$). The situation, for a
couple of cases, is illustrated in Fig. 1, where is plotted the function $(F-1)$, which
coincides with $2V$ when $|F-1| \simeq 0$.\\
\begin{figure}[h]
\epsfxsize=7.5cm
\epsfbox{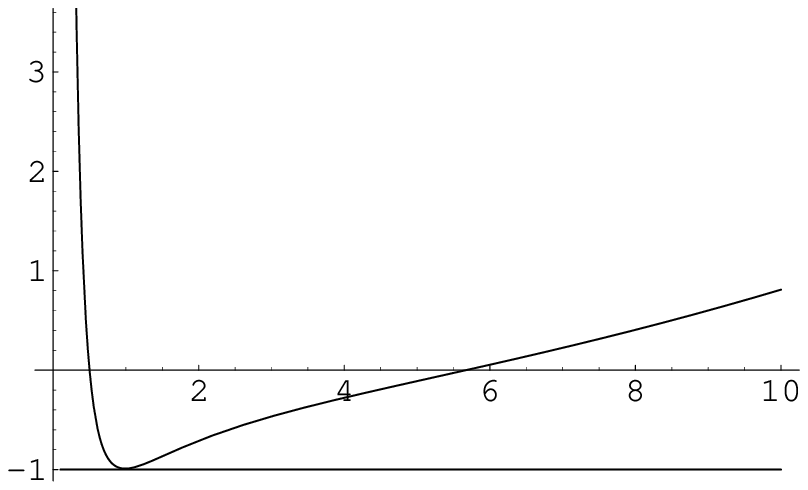}
\epsfxsize=7.5cm
\epsfbox{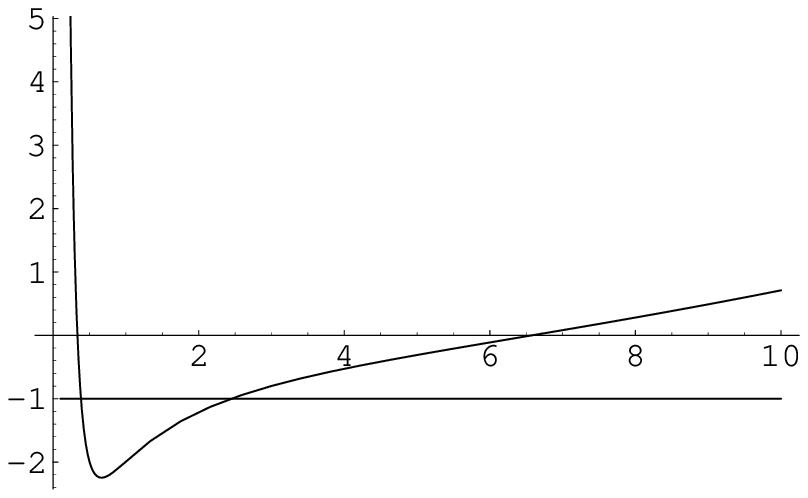}
%
\caption{For $k=1$, $\lambda > 0$, the cases of two coincident horizons and two
different horizons.}
\end{figure}
\\
In the diagrams, the horizons are located at the intersection point(s)
between the graph of $F-1$ and the straight line $y=-1$. The regions
where $V$ can be defined, and where $2V \simeq (F-1)$, are located in the
neighborhood of the (second) intersection point between the graph of
$F-1$ and the axis $y=0$.
When $\lambda \to 0^{+}$ (small cosmological constant) and $Q^2 \to 0$
(small electric charge of the hole), the zero going to the Schwarzschild
horizon, $R=2M$, is the outer one (see also Ref.~\cite{louko}).\\
\\
For $k=1$ and $\lambda <0$ (de Sitter space-time) the situation is a bit more
complicated. The analysis reveals that there
is always one negative non physical zero. As regard the positive zeros of
$F(r)$ (i.e. those originating the horizons), we can have just one zero,
2 coincident zeros and another positive zero, 3 different zeros, one zero
and other two coincident zeros, one zero only. The second and third of these
circumstances are
summarized in the diagrams of Fig. 2, where again the function $(F-1)$ is plotted.
As before, $V$ is defined in the regions where $|F-1|\simeq0$.
The third case, the one with three positive zeros, is probably the most
representative from the physical point of view, with $M\gg|Q|\gg|\lambda|$.
In this case, the zero tending to the Schwarzschild value, $R \to 2M$,
when $\lambda \to 0^{-}$ and $Q^2\to 0$, is the "middle" one.
\begin{figure}[ht]
\epsfxsize=7.5cm
\epsfbox{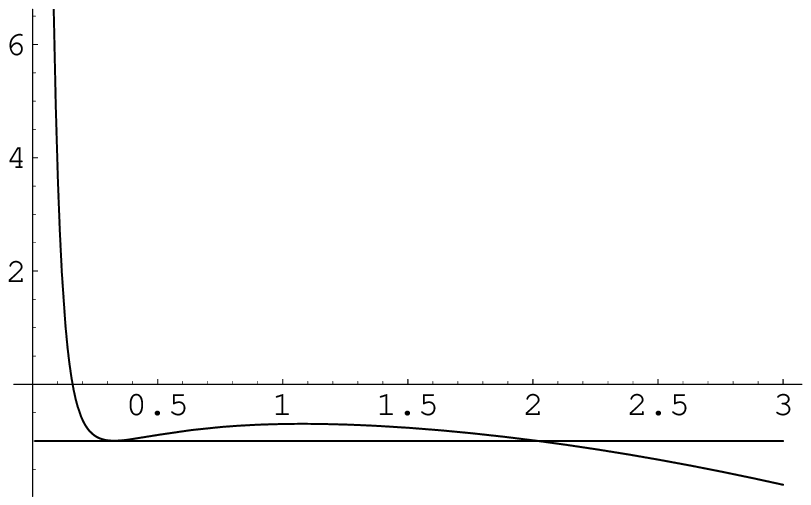}
\epsfxsize=7.5cm
\epsfbox{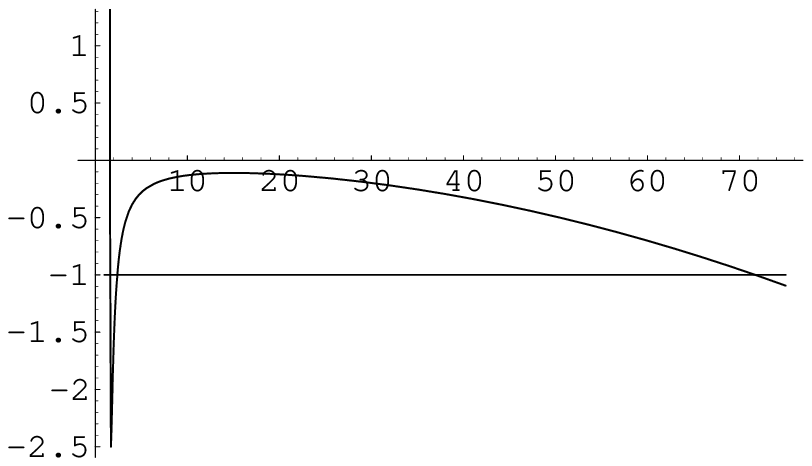}
%
\caption{For $k=1$, $\lambda<0$, the cases of 2 coincident plus 1
horizons, and of 3 different horizons.}
\end{figure}
\\
Analog diagrams can be drawn for the situations $k=0$ ($\lambda > 0,\, \lambda< 0$), and
$k=-1$ ($\lambda > 0,\, \lambda< 0$). From them we can visualize, as before, the regions
where the effective potential $V$ can be defined.\\
A very useful parametrization for the black hole space-times, instead of
the pair $(M,\,Q)$, is the pair
$(R_h,\,Q)$, where $R_h$ is the value of $r$ at the (outer)(or middle)
horizon. The mass is then given
(from the eq. $F(R_h)=0$) in terms of $(R_h,\,Q)$ as
\be
M=\frac{R_h}{2}\left(\lambda R_h^2 + k + \frac{Q^2}{R_h^2}\right).
\label{mass}
\ee
As we see, the relation (\ref{mass}) yields the usual Schwarzschild
relation (namely $R_h=2M$) for $\lambda \to 0$, $Q \to 0$, $k=1$.
\section{Hawking temperature from Heisenberg uncertainty principle}
\setcounter{equation}{0}
In order to get the Hawking formula, consider the following gedanken
experiment.
Suppose to have a gravitational field described by the metric
(\ref{mo})(\ref{m}) and a region where $|F-k|\simeq 0$.
There the effective potential $V$ can be defined, and the eq. (\ref{vf}) holds.
Imagine to have a neutral (uncharged) particle of rest mass $m$
which falls radially in such a field. Then
the newtonian potential energy of the particle, due to gravity,
is, classically,
\be
U=mV=\frac{1}{2}\,mc^2(F-k).
\label{pe}
\ee
Let us now suppose that this expression of the classical potential energy holds in any region,
also where $|F-k|$ differs strongly from zero. This means to extrapolate the validity of
eq. (\ref{pe}), which is a weak field result, to a strong field situation.
If the particle falls for a small radial displacement $\Delta r$, then
it feels a variation in potential energy equal to
\be
\Delta U = \frac{1}{2}\,mc^2\, F'(r)\,\Delta r .
\ee
During the shift $\Delta r$, the particle acquires a kinetic energy $\Delta K$
which equals the lost potential energy $\Delta U$. Suppose that $\Delta K$
is sufficient to create a particle-antiparticle pair from the quantum vacuum,
\be
\Delta K = 2 mc^2 .
\ee
Then we can compute the $\Delta r$ needed for this process
\be
\Delta U = \Delta K = 2 mc^2 \quad \Rightarrow \quad F'(r)\Delta r = 4\,,
\ee
that is
\be
\Delta r=\frac{4}{F'(r)}.
\ee
This process, repeated many times, creates a gas of particles in the
region where it takes place. Since the particles are confined in a space
slice of thickness $\Delta r$, each of them has an uncertainty in
(kinetic) energy equal to
\be
\Delta E = \frac{\hbar c}{2 \Delta r} = \frac{\hbar c}{8}\,F'(r).
\ee
Suppose now to interpret the quantum uncertainty in the (kinetic) energy
of these particles as due to thermal agitation, which means that we can
write, again using simply the classical Maxwell-Boltzmann statistics,
\be
\Delta E \sim \frac{3}{2}k_B T
\ee
where $T$ is the temperature of this gas of particles. Therefore
\be
\frac{3}{2}k_B T = \frac{\hbar c}{8}\,F'(r)
\ee
or
\be
T = \frac{\hbar c}{12\, k_B}\,F'(r).
\label{T}
\ee
This is the temperature of the particle gas confined in the space
slice $\Delta r$ around the value $r$ of the radial coordinate.\\
Imagine that the whole process just described takes place close
to the external edge of the (event) horizon, i.e. for $r\simeq R_h$.
By the hypothesis assumed, the expression (\ref{pe}) for the effective potential energy $U$
maintains its validity even close to the outer (or middle) horizon, i.e. for $r\simeq R_h$.
In other words, we extrapolate the validity of the expression (\ref{pe}) for
$U$ even to the region $r\simeq R_h$
\footnote{We know from the previous sections that the effective
potential $V$ (because of the hypothesis $|h_{00}|\ll 1$) can be defined
only in regions where $|F-k|\simeq0$. In such regions we have
$V=(F-k)c^2/2$ (eq. (\ref{vf})). On the contrary, in the proximity of
an (event) horizon, $r\simeq R_h$, we have by definition $F\simeq0$,
that is $(F-k)\simeq k$. Therefore, in such regions, the effective
potential could not, rigorously speaking, be defined.}.
This will be sufficient
to get the Hawking temperature of the hole with a very good agreement
with the QFT formula.\\
In fact, from eqs. (\ref{T}) and (\ref{m}) we can write for $r=R_h$
\be
T=\frac{\hbar c}{12\, k_B}\left(\frac{2GM}{c^2R_h^2}-\frac{2GQ^2}{c^4R_h^3}+
2\lambda R_h\right)
\ee
and, inserting $M$ from the mass formula (\ref{mass}) with the constants restored
\be
\frac{2GM}{c^2} = kR_h + \frac{GQ^2}{c^4 R_h} + \lambda R_h^3 \,,
\ee
we have finally
\be
T=\frac{\hbar c}{12\, k_B}\left(\frac{k}{R_h}-\frac{GQ^2}{c^4R_h^3}+3\lambda R_h\right).
\label{TT}
\ee
This is the temperature of the particle gas thickened on the surface
of the (event) horizon, that is the temperature of the hole, as seen
by an observer set far in the quasi-flat region of the space-time.
In particular, for a Schwarzschild black hole we can write
\be
T=\frac{\hbar c}{12\, k_B\, R_h}=\frac{\hbar c^3}{24\, k_B \,GM}.
\ee
It is remarkable that the exact formula for the Hawking temperature,
namely
\be
T=\frac{\hbar c}{4\pi\, k_B}\left(\frac{k}{R_h}-\frac{GQ^2}{c^4R_h^3}+
3\lambda R_h\right)\,,
\ee
worked out with the apparatus of QFT in curved space-time (see Ref.~\cite{louko}),
differs from the (\ref{TT})
only for the factor $4\pi \simeq 12$. This is especially remarkable if we think
that for the derivation of (\ref{TT}) we used only laws of classical
physics plus the Heisenberg principle.
\section{Comment and conclusion}
\setcounter{equation}{0}
We note that this derivation can be used for the temperature of the horizons of acoustic
black holes and, due to its generality, wherever a metric admitting an effective
(newtonian) potential is present. So, in this paper we have shown that it is possible to
compute the Hawking temperature for a large class of black holes (Schwarzschild,
Reissner-Nordstr\"{o}m, (Anti) de Sitter) in a unified way, using always only the
Heisenberg uncertainty principle, plus laws from classical physics. The Hawking
temperature does not seem, also in this analysis, to have any link with the dynamical
Einstein equations. Moreover, no generalized uncertainty principle has been used, in
any form. As a byproduct we can therefore say that the fact that (A)dS black hole
thermodynamics admits a holographic description in terms of dual conformal QFT, whereas
Schwarzschild black hole does not, does not seem to rely or depend on the diversity of
the uncertainty principle(s) used to derive such thermodynamics.
\section*{Acknowledgement}
The author wishes to thank J.P.S. Lemos for having drawn his attention on the references
\cite{Cavaglia}, \cite{louko}.
%
%

%

\begin{thebibliography}{99}
%
%
\bibitem{Scardigli}
F.~Scardigli, Nuovo Cimento {\bf B 110}, 1029 (1995) [arXiv: gr-qc/0206025].
%
%
\bibitem{Santiago}
R.J.~Adler, P.~Chen, D.I.~Santiago, Gen. Rel. Grav. {\bf 33}, 2101 (2001)
[arXiv: gr-qc/0106080].
%
\bibitem{Cavaglia}
B.~Bolen, M.~Cavaglia, Gen. Rel. Grav. {\bf 37}, 1255 (2005) [arXiv: gr-qc/0411086].
%
\bibitem{louko}
D.R.~Brill, J.~Louko, P.~Peldan, Phys. Rev. {\bf D 56}, 3600 (1997) [arXiv: gr-qc/9705012].
%
%
\bibitem{Visser}
M.~Visser, Int. J. Mod. Phys. {\bf D 12}, 649 (2003) [arXiv: hep-th/0106111].
%
%
\bibitem{weinberg}
S.~Weinberg, {\em Gravitation and Cosmology: principles and applications of the
General Theory of Relativity}, J.~Wiley \& Sons, New York, 1972.
%
%
\bibitem{43louko}
M.~Walker, J. Math. Phys. (N.Y.) {\bf 11}, 2280 (1970).
%
%
\bibitem{71030louko}
K.~Lake, Phys. Rev. {\bf D 19}, 421 (1979).\\
M.~Banados, C.~Teitelboim, J.~Zanelli, Phys. Rev. {\bf D 49}, 975 (1994).\\
J.~Louko, S.N.~Winters-Hilt, Phys. Rev. {\bf D 54}, 2647 (1996).
%
%
%
%
\end{thebibliography}
\end{document}